\input{aipcheck}
\documentclass[final]{aipproc}
\layoutstyle{6x9}

\begin{document}

\title{Radiative Processes\\ In Extragalactic Large-Scale Jets}


\author{\L ukasz Stawarz}{
address={Max-Planck-Institut fur Kernphysik, PO Box 103980, 69029 Heidelberg, Germany\\ \texttt{Lukasz.Stawarz@mpi-hd.mpg.de}\\ Obserwatorium Astronomiczne, Uniwersytet Jagiello\'nski, ul. Orla 171, 30-244 Krak\'ow, Poland}
}

\begin{abstract}
About one thousand extragalactic large-scale jets are known, and a few tens of them are confirmed sources of infrared, optical, or X-ray photons. Multiwavelength emission comming directly from these outflows is always non-thermal in origin. This fact constitutes a primary difficulty in extracting unknown parameters of large-scale jets, since the non-thermal featureless continua do not allow to infer undoubtfully (or even at all) bulk velocities and composition of the radiating plasma. In addition, arcsecond spatial resolution, limited sensitivity and narrow energy bands of the best high-frequency telescopes like Spitzer, Hubble and Chandra, preclude precise constraints on the spectral and morphological properties of the discussed objects. Nevertheless, new multiwavelength observations have substantially enriched our knowledge on extragalactic large-scale jets, in many aspects, however, by means of challenging previous predictions and expectations. In this short contribution I will concentrate on the following issue: what can be learned by analyzing broad-band emission of the discussed objects about particle acceleration processes acting thereby and about jet internal parameters.
\end{abstract}

\maketitle

\section{Particle Acceleration Processes}

Polarization and power-law spectral distribution of radio emission observed from extragalactic large-scale jets imply its synchrotron origin, and thus a power-law energy distribution of the radiating ultrarelativistic electrons. Radio spectral indices are concentrated around $\alpha_{\rm R} \sim 0.75$ (see, e.g., Kataoka \& Stawarz 2005). This value is not strictly predicted by any particular model of particle acceleration applied to the discussed objects. For example, assuming the `universal' shock-like particle spectrum $n_{\rm e}(\gamma) \propto \gamma^{-2}$ usually considered in this context, where $\gamma$ is the particle Lorentz factor, one expects the synchrotron continuum of the form $S_{\nu} \propto \nu^{-\alpha}$ with the spectral index $\alpha = 0.5$ in a weak cooling regime, and $\alpha = 1.0$ in a strong synchrotron (expected to be dominant) cooling regime. The situation is therefore that even the observed radio spectral properties of the large-scale jet are not consistent with the simplest scenarios for the evolution of the electron energy distribution. Taking into account the complexity of the jet phenomenon we should not in fact expect the simplest models to be realistic. Meanwhile, the `injection' electron spectrum (not affected by spectral ageing) remains elusive. One can hope that by performing observations at low radio frequencies such a spectrum will be revealed, and thus strong observational constraints will help to develop the appropriate theoretical model. Indeed, Young et al. (2005) found that the radio data for FR I sources imply $n_{\rm e}(\gamma) \propto \gamma^{-2.1}$. Such a spectrum may be interpreted as manifestation of the effects connected with relativistic velocities of the internal shocks accelerating jet particles (see, e.g., Kirk \& Duffy 1999). On the other hand, this conclusion is still premature, keeping in mind recently discussed problems with the `first order Fermi' shock acceleration in the relativistic regime (e.g., Niemiec \& Ostrowski 2004).

\begin{figure}
  \includegraphics[height=.3\textheight]{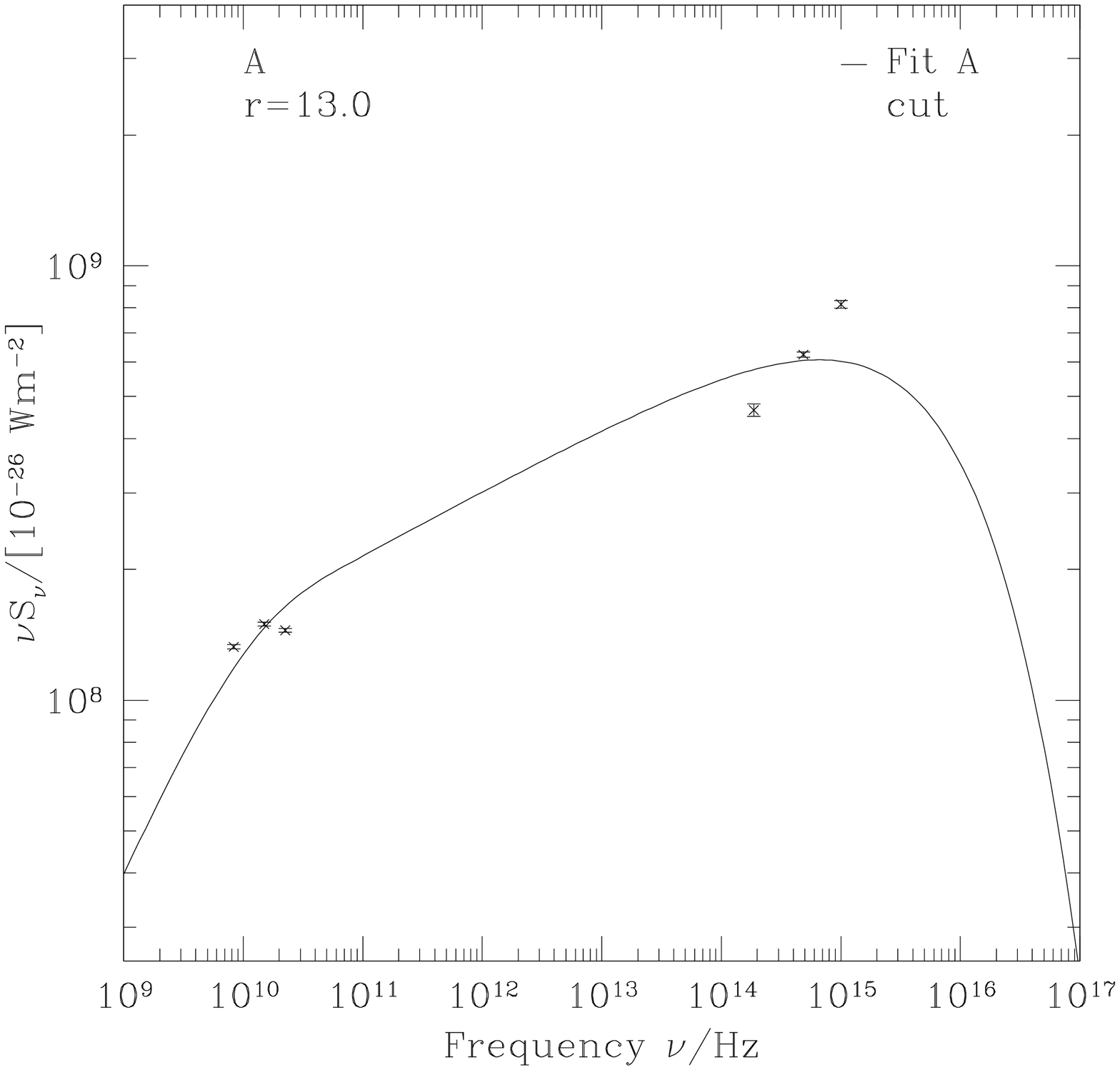}
  \includegraphics[height=.32\textheight]{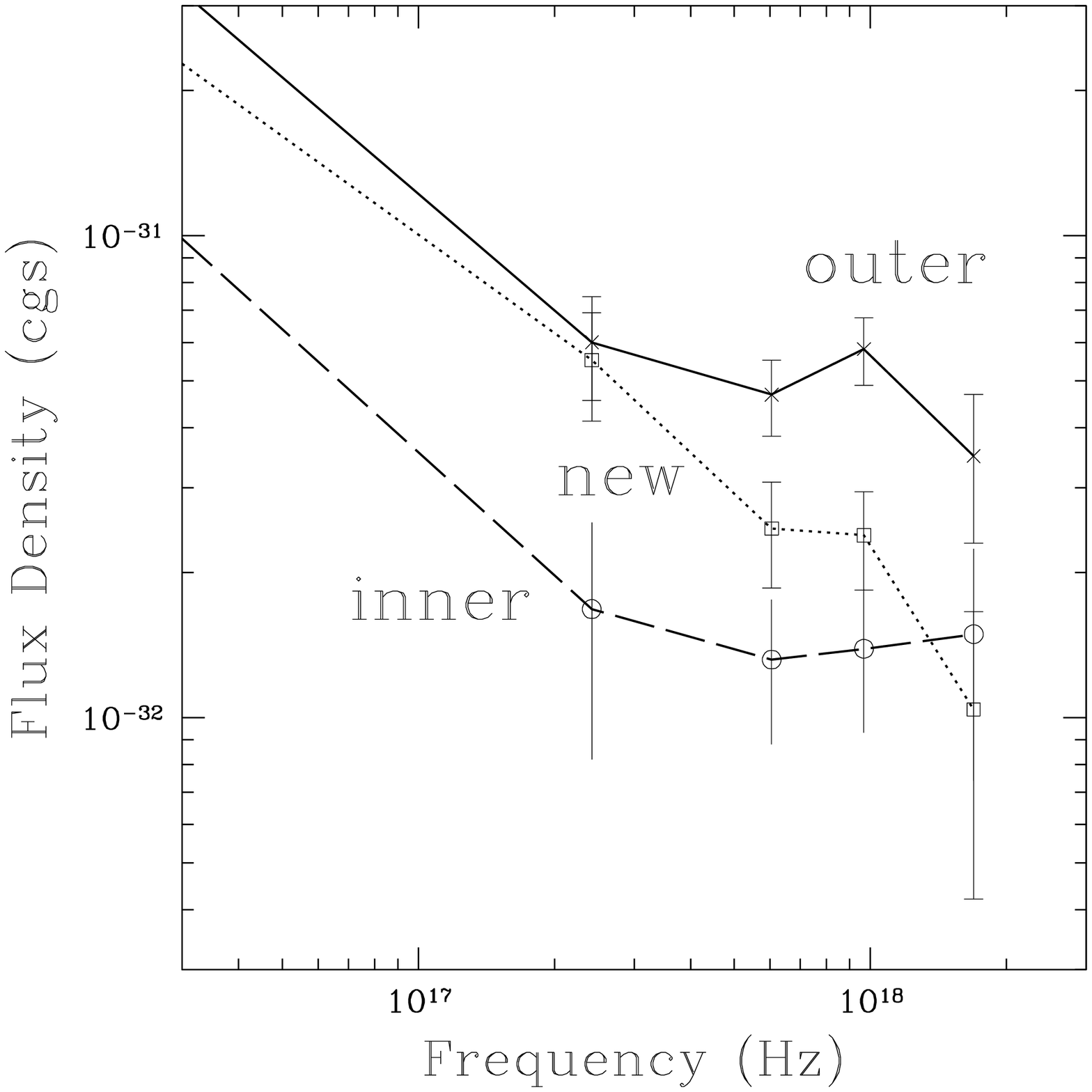}
  \caption{{\it Left:} Knot A in 3C 273 jet in radio and optical frequencies (from Jester et al. 2005). {\it Right:} Different components of the knot K25 in 3C 120 jet in X-ray frequencies (from Harris et al. 2004).}
\end{figure}

Polarized optical emission of extragalactic large-scale jets is generally believed to constitute a high energy part of the synchrotron continuum. With the equipartition jet magnetic field $B_{\rm eq} \sim 10 - 100$ $\mu$G and moderate beaming, the Lorentz factor of the electrons emitting optical synchrotron photons is roughly $\gamma \sim 10^6$. Thus, any process for the particle acceleration applied to the discussed objects has to be able to produce electrons with TeV energies. Is it however the maximum? The answer is no, since apart from convincing evidence for the spectral cut-off occurring at the observed frequencies $10^{14}-10^{15}$ Hz, there are also cases with smooth continuation of the optical fluxes up to higher photon energies. We note in this context a large scatter of the optical spectral indices, observed in the range $\alpha_{\rm O} \sim 0.5-2.0$. In fact, one should expect significant spectral ageing of the high-energy electrons, leading to an increase of the observed spectral index and a decrease of the maximum synchrotron frequency away from the particle acceleration sites. However, the most recent observations indicate that in some cases the optical spectra are flatter than the radio spectra, and that the steepening of the optical continua along the jets is generally too slow. An example of unexpected optical behavior --- the brightest knot A of 3C 273 jet (see Jester et al. 2005) --- is shown on figure 1. Spectral energy distribution of this knot challenges usually considered homogeneous one-zone models for the knots in extragalactic large-scale jets, and modeling of their spectra in terms of a single, or eventually smoothly broken power-law electron energy distribution. Instead, non-homogeneous emission sites and non-standard electron spectra have to be seriously taken into account.

Optical observations put also another important constraint on the particle acceleration processes acting in extragalactic large-scale jets: they indicate a need for a distributed, continuous energetization of the radiating electrons all along the outflow, not limited exclusively to its brightest parts, i.e. to the knots (which are usually identified with the extended shock waves). The reason is that we observe synchrotron optical radiation also from the interknot regions, for which the extension is much larger than that allowed by the electrons' radiative cooling time scale (see Stawarz 2003 for a more detailed discussion). As shown by Jester et al. (2001) for the 3C 273 jet, this discrepancy cannot be removed simply by changing some jet parameters like the jet bulk Lorentz factor or the magnetic field intensity. Let us only mention in this context, that there is growing evidence that an issue of continuous particle acceleration is of significant importance in high energy astrophysics. Different physical processes can account for energetization of ultrarelativistic particles throughout the whole volume of the extended astrophysical sources, among which stochastic interaction of relativistic particles with MHD/plasma waves and turbulence is the best understood at the moment. This kind of acceleration process was therefore applied to explain non-thermal emission of many different objects, including solar loops (Petrosian \& Donaghy 1999), accreting plasma in the Galactic supermassive black hole Sgr A$^*$ (Liu et al. 2004), galaxy clusters (Brunetti et al. 2004), or finally extragalactic jets themselves (e.g., Stawarz \& Ostrowski 2002).

X-ray observations showed that the large-scale jets are surprisingly bright in X-rays. In the case of FR I sources, the observed X-ray spectral indices of the knots are steep, $\alpha_{\rm X} \geq 1.0$, and their radio-to-X-ray continua (peaking usually around infrared frequencies) can be well fitted by broken power-laws, indicating synchrotron origin of the keV photons (e.g., Hardcastle et al. 2001 for 3C 66B jet). Hence, the maximum electron energies are pushed up to $10-100$ TeV range (electron Lorentz factors $\gamma \sim 10^7 - 10^8$). In fact, the ability to accelerate ultrarelativistic electrons to such high energies seems to be a general property of turbulent astrophysical outflows, since synchrotron keV photons are also detected from a number of supernova remnants (see Bamba et al. 2005) and microquasar jets (e.g., Corbel et al. 2005). We note, that in all of the discussed objects the equipartition magnetic field is roughly in a range $10-100$ $\mu$G. The synchrotron scenario for the X-ray emission can be as well applied to powerful large-scale quasar jets. In this case, however, non-standard deviations from a single power-law energy distribution of the radiating electrons --- in a sense of high-energy spectral hardening --- are required, since the X-ray emission of these objects is usually much higher than implied by extrapolation of radio-to-optical continua (e.g., Schwartz et al. 2000 for PKS 0637-752 jet). We note, that the non-standard X-ray spectra are observed directly in the 3C 120 jet (Harris et al. 2004 and figure 1). Indeed, the required spectral hardening is expected if the effective particle acceleration acts continuously within the extended emission region (Stawarz \& Ostrowski 2002), or if the electrons cool predominantly due to the inverse-Compton emission in the Klein-Nishina regime (Dermer \& Atoyan 2002). Alternatively, additional high energy flat-spectrum electron or proton population can be considered (Aharonian 2002). Instead, the currently favored model involves inverse-Compton emission of low-energy electrons ($\gamma \sim 100$) on the CMB radiation field (Tavecchio et al. 2000, Celotti et al. 2001). This model, requiring large bulk Lorentz factors of the jets ($\Gamma > 10$) on hundreds of kpc scales, implies also that the electron energy distribution has to continue down and cut-off sharply at $\gamma \sim 10$ in order not to overproduce optical fluxes of the knots.

\begin{figure}
  \includegraphics[angle=90,height=.29\textheight]{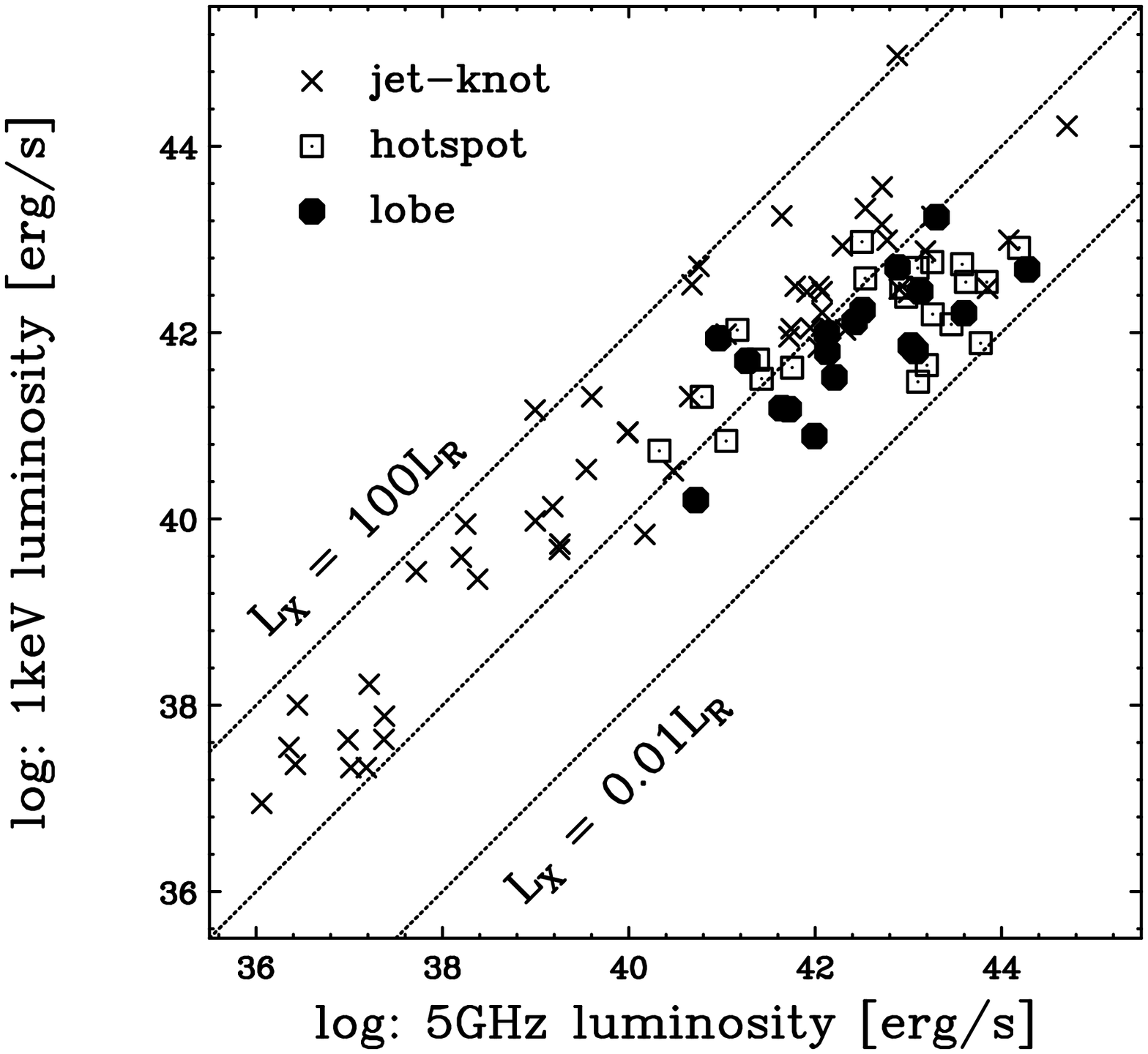}
  \hspace{0.3in}
  \includegraphics[height=.3\textheight]{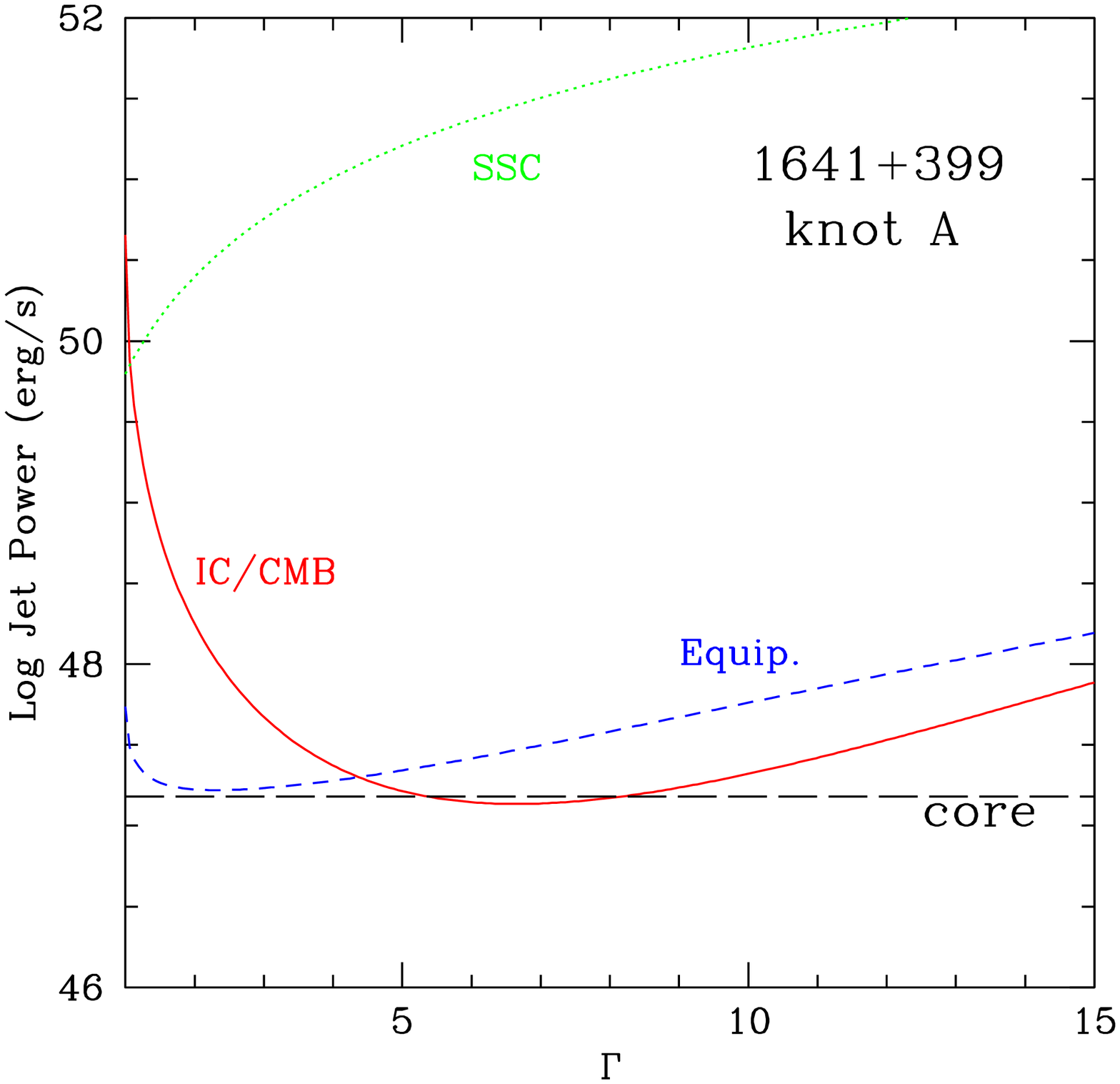}
  \caption{{\it Left:} The observed range of $1$ GHz and $1$ keV luminosities for the large-scale jets, lobes and hotspots of 44 radio sources (from Kataoka \& Stawarz 2005). {\it Right:} Energetics of the jet in quasar 1641+399 infered from the IC/CMB and synchrotron self-Compton models for the X-ray emission of its outer parts, from the equipartition condition, and from modeling of its blazar core, as indicated near each curve (from Tavecchio et al. 2004).}
\end{figure}

\section{Jet Internal Parameters}

Radio and X-ray luminosities of extragalactic large-scale jets range from $L \sim 10^{36}$ erg s$^{-1}$ in the case of the weakest FR I sources up to $10^{45}$ erg s$^{-1}$ for the most powerful quasars (see Kataoka \& Stawarz 2005 and figure 2). Thus, the radiative output is orders of magnitude less than total kinetic power (inferred from, e.g., lobes' energetics) of the discussed objects. Interestingly, in a framework of the IC/CMB model the total kinetic power of the quasar jets has to be constant along the outflow from sub-pc up to hundreds of kpc-scale (see Tavecchio et al. 2004 and figure 2). This implies that the powerful jets are inefficient radiators, but instead have to efficiently transport almost all the energy in the form of relativistic bulk motion of cold particles without any losses up to Mpc distances from the active core (see in this context Gallo et al. 2005 for the jet in microquasar Cygnus X-1). Another question is if this power is transported at a constant rate during the whole jet activity epoch, or not. Stawarz et al. (2004) proposed that the observed morphological properties of large-scale quasar jets (in particular, frequency-independent knots' profiles) are most likely manifestation of a modulated jet activity. In this picture, considered for some time before (Reynolds \& Begelman 1997) and constituting nice connection to the jet activity observed in X-ray binary systems, knots represent portions of the jet matter with excess kinetic power, resulting from $10^4$-year-long high activity periods separated by $10^5$-year-long epochs of lower radio activity. Let us mention, that such an interpretation is independent on the particular process responsible for production of the X-ray emission of large-scale quasar jets. Although, as shown recently by Uchiyama et al. (2005) for the PKS 0637-752 jet, this is required if the IC/CMB model is the case.

\begin{figure}
  \includegraphics[height=.28\textheight]{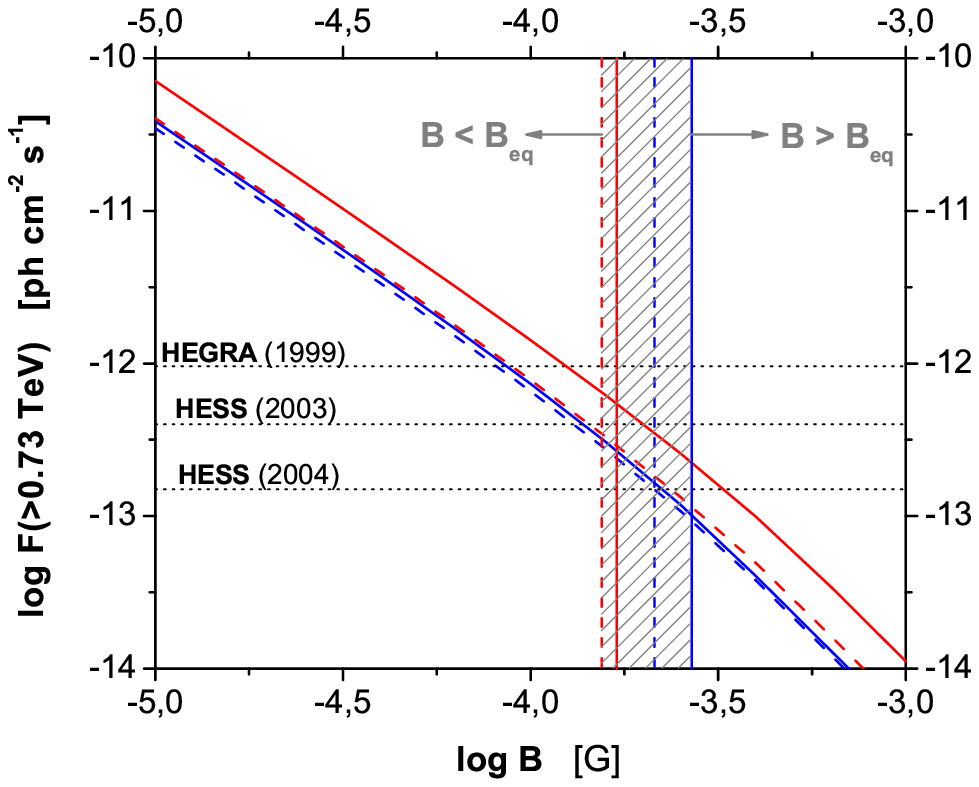}
  \includegraphics[height=.28\textheight]{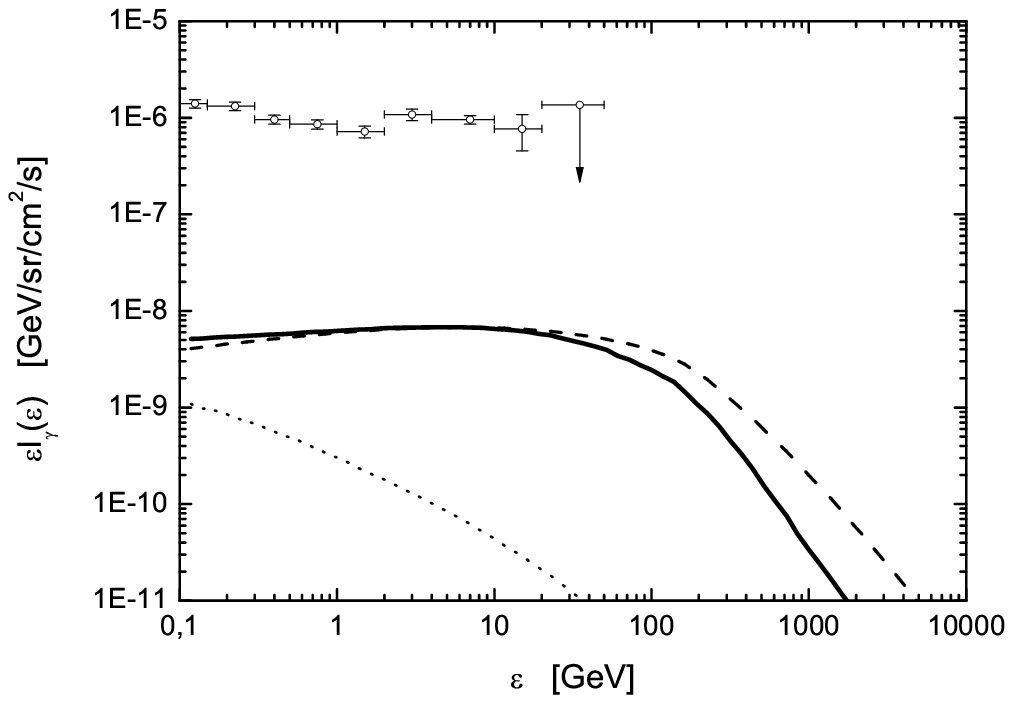}
  \caption{{\it Left:} Constraints on the jet magnetic field within knot A of M 87 jet imposed by the HEGRA and HESS observations (dotted horizontal lines) for different possible jet parameters. Vertical lines denote the equipartition magnetic field (Stawarz et al. 2005). {\it Right:} Contribution of FR~I kpc-scale jets to the extragalactic $\gamma$-ray background for. Open circles correspond to the extragalactic EGRET $\gamma$-ray background. Dashed lines indicate intrinsic emission, thick solid lines correspond to the emission with absorption/re-emission effects included, while dotted lines illustrate emission from the source's halo (from Stawarz et al. 2006).}
\end{figure}

The intensity and structure of the magnetic field is an another open problem in understanding extragalactic large-scale jets. Even though most of the models for the jet launching in AGNs involves a dynamically dominating magnetic field, it is not clear if such dominance holds also at the larger scales, and, if not, where the conversion of the Poynting flux to the bulk kinetic power of the jet particles is taking place (see a discussion in Sikora et al. 2005). Again, analysis of the broad-band emission of these outflows on kpc and hundreds of kpc-scales may help to answer at least the first part of this question by constraining the jet content. For example, if the IC/CMB hypothesis for the quasar jet X-ray emission is correct, one can say that it is possible to find some particular value of a bulk Doppler factor which allows for the energy equipartition between the jet magnetic field and the jet particles (of electron-proton content; see Ghisellini \& Celotti 2001). Yet, the required in this way values for the jet bulk Lorentz factors are in some cases uncomfortably large, and also the assumed in this approach homogeneity of the kpc-scale emission regions is questionable (see a discussion in Kataoka \& Stawarz 2005).

In the case of the low-power FR I jets the situation is however different. The established synchrotron origin of the jet keV photons, kpc-distances from centers of host galaxies, expected sub- or only mildly relativistic bulk velocities, and well-covered (in some cases) synchrotron continua of these outflows, allow to analyze relatively precisely their expected very high energy $\gamma$-ray emission resulting from inverse-Compton scattering of the ambient photon fields (especially the dominant starlight emission of the elliptical hosts; Stawarz et al. 2003). The predicted fluxes, as functions of the unknown jet magnetic field, can be then compared in the case of the most nearby objects with the observed upper limits provided by modern Cheronkov Telescopes, obtaining some meaningful constraints. For example, Stawarz et al. (2005) showed that for the brightest knot A in M 87 jet, recent HEGRA and HESS observations (Beilicke et al. 2005) imply that the magnetic field intensity thereby cannot be smaller than the equipartition value $B_{\rm eq} \sim 300$ $\mu$G, as otherwise the observed $\gamma$-ray flux would be overproduced. For the whole class of FR I jets, Stawarz et al. (2006) obtained somewhat less strong constraints by means of analyzing the expected contribution of these objects to the extragalactic $\gamma$-ray background as measured by EGRET. In particular, Stawarz et al. (2006) found that with the equipartition magnetic field $100$ $\mu$G the appropriate contribution from the FR I kpc-scale jets is about $1\%$. As the expected $\gamma$-ray flux scales roughly with the square of the jet magnetic field intensity, this result indicates a safe lower limit $B > 10$ $\mu$G, because for a lower $B$ the diffuse $\gamma$-ray background would be overproduced. 

\begin{theacknowledgments}
It is a pleasure to thank all my jet colleagues and collaborators. This work was supported by the grant PBZ-KBN-054/P03/2001.
\end{theacknowledgments}

\end{document}